# Nanometer Scale Spectral Imaging of Quantum Emitters in Nanowires and its correlation to their atomically resolved structure

Luiz Fernando Zagonel[1], Stefano Mazzucco[1], Marcel Tencé[1], Katia March[1], Romain Bernard[1], Benoît Laslier[1], Gwénolé Jacopin[2], Maria Tchernycheva[2], Lorenzo Rigutti[2], Francois H. Julien[2], Rudeesun Songmuang[3], Mathieu Kociak[1,*]

[1]Laboratoire de Physique des Solides, Bâtiment 510, CNRS UMR 8502, Univ. Paris Sud XI, F 91405 Orsay, France

[2]Institut d'Electronique Fondamentale, CNRS UMR8622, Univ. Paris-Sud, 91405, Orsay, France [3]CEA-CNRS group "Nanophysique et Semiconducteurs", Institut Neel, Grenoble, France

[*]To whom correspondence should be addressed; E-mail: kociak@lps.u-psud.fr

ABSTRACT

We report the spectral imaging in the UV to visible range with nanometer scale resolu-tion of closely packed GaN/AlN quantum discs in individual nanowires using an improved custom-made cathodoluminescence system. We demonstrate the possibility to measure full spectral features of individual quantum emitters as small as one nanometer and separated from each others by only few nanometers, and the ability to correlate their optical properties to their size, measured with atomic resolution. The direct correlation between the quantum disc size and emission wavelength allows us to evidence the quantum confined Stark effect leading to an emission below the bulk GaN band gap for discs thicker than 2.6 nm. Helped with simula-tions, we show that the internal electric field in the studied quantum discs is smaller than what is expected in the quantum well case. We evidence a clear dispersion of the emission wave-lengths of different quantum discs of identical size but different position along the wire. This dispersion is systematically correlated to a change of the diameter of the AlN shell coating the wire, and is thus attributed to the related strain variations along the wire. The present work opens the way both for fundamental studies of quantum confinement in closely packed quan-tum emitters and for characterizations of optoelectronic devices presenting carrier localization on the nanometer scale.

Quantum confined systems such as quantum discs (QDiscs) and quantum dots (QD) are arti-ficial systems made up of a low energy gap semi-conductor embedded in a larger gap one, which mimics the behaviour of atoms with on-purpose designs. [1,2] These systems are heavily used in many applications, ranging from lasers, [3] photodetectors [4] to quantum bits or single photon sources, [5] where quantum confinement is exploited for tailoring the electronic and optical properties of devices. Although simple in its principle, the quantum confinement of electrons in artificial systems triggers both theoretical and practical questions. Indeed, the excitation spectra of quantum emitters (QE) depend on their size, generally only a few atomic layers, and can also be strongly affected by variations of their electronic and electromagnetic environment at the nanometer scale. In the special case of GaN/AlN QE which are widely used for their optical properties in the visible-UV domain, a size change by only a single atomic layer leads to drastic change of the electronic and optical properties. [6] On a larger, nanometric to micrometric scale, the presence of electrical fields, should they be intrinsic (pyro- or piezoelectricity) or extrinsic will drastically change the properties of the QE. [7,8] The combination of a very high spatial and spectral resolution required to study single quantum emitters in an arbitrary environment and to correlate the emission spectrum to their sizes measured at the monolayer (ML) resolution has never been achieved before, despite a tremendous effort in that direction. Photoluminescence (PL) experiments are widely used to study QEs, from large populations [8] to small ensembles (as they appear in individual nanowires, for example). [9] However, due to the limited spatial resolution (about half the wavelength at best), uncertainties remain on the attribution of optical transitions to the individual QEs. It seems thus interesting to switch to the alternative given by the optical spectroscopy with fast electrons in electron microscopes. Indeed, it has recently been proved that Electron Energy Loss Spectroscopy (EELS) in a Scanning Transmission Electron Microscope (STEM) made possible to image optical properties of metallic nanoparticles with few nanometers resolution. [10] However, the EELS energy resolution (routinely 300 meV and at the very best 100 meV) is typically one order of magnitude worse than needed. Moreover, the EELS signal is not related to the emission, but rather to the absorption of the system. Cathodoluminescence (CL), on the opposite, has been used with success in studying luminescence properties of semiconductors [11-14] and plasmons [15] with a much better spectral resolution. However, the state-of-the-art cathodoluminescence [11-14,16,17] gives the required spectral information but cannot resolve closely packed QEs. In the case where the signal at a single wavelength is recorded (and thus most of the spectroscopic interest is lost), the best spatial reported resolutions were around 10 nm [16] to 20 nm [17] (compared to a more routine value of 50 nm [12]). The best achieved resolution was therefore one order of magnitude too large to probe optical spectroscopic information of individual QEs in many real devices, with the parallel determination of the structure with atomic resolution. Moreover, the insight gained in full spectra was lost in those cases, and only intensity maps could be gathered - the single QE limit was thus not reachable. This triggers the need for routine optical spectroscopic techniques with spatial resolution better than the typical size of a QE (i.e. 1 to 5 nm) which could be correlated with atomically resolved structural information.

In this letter, we report the spectral imaging in the UV to visible range with nanometer scale resolution of tightly packed QEs using an improved custom-made CL system. We demonstrate the measurement of full spectral features of individual QEs as small as one nanometer and separated from each other by only few nanometers, and the ability to correlate their optical properties to their size, measured with atomic resolution. The QEs under study belong to a stack of GaN/AlN QDiscs in a single nanowire. The 2D mapping of the full spectroscopic information allows us to clearly attribute emission lines to the QDiscs, ie excluding the possibility of emission from defects or small QEs in the sides of the NWs. We directly evidence a strong redshift of the emission energy with increasing QDisc size as a consequence of quantum confined Stark effect. Still, the compar-ison between the cathodoluminescence results and the modeling performed with experimentally measured QDisc sizes shows that the internal electric field in the QDiscs is reduced with respect to the quantum well case. The influence of strain exercised by a lateral AlN shell on the peak emission wavelength has been demonstrated as well as the impact of the shell on the luminescence intensity. In particular, we evidence the effect of the strain variation on the energy dispersion be-tween QDiscs of identical size. The spectral resolution of the present technique was as high as a few tens of meV. These unprecedented combined spatial and spectral resolutions should benefit to the study of many optoelectronic structures such as Stranski-Krastanow quantum dots or quantum wells presenting in-plane carrier localization. [18-20] The unique possibility to correlate optical and structural properties on the same nanoobject will also help to refine material parameters and serve to validate current quantum confinement theories. [21]

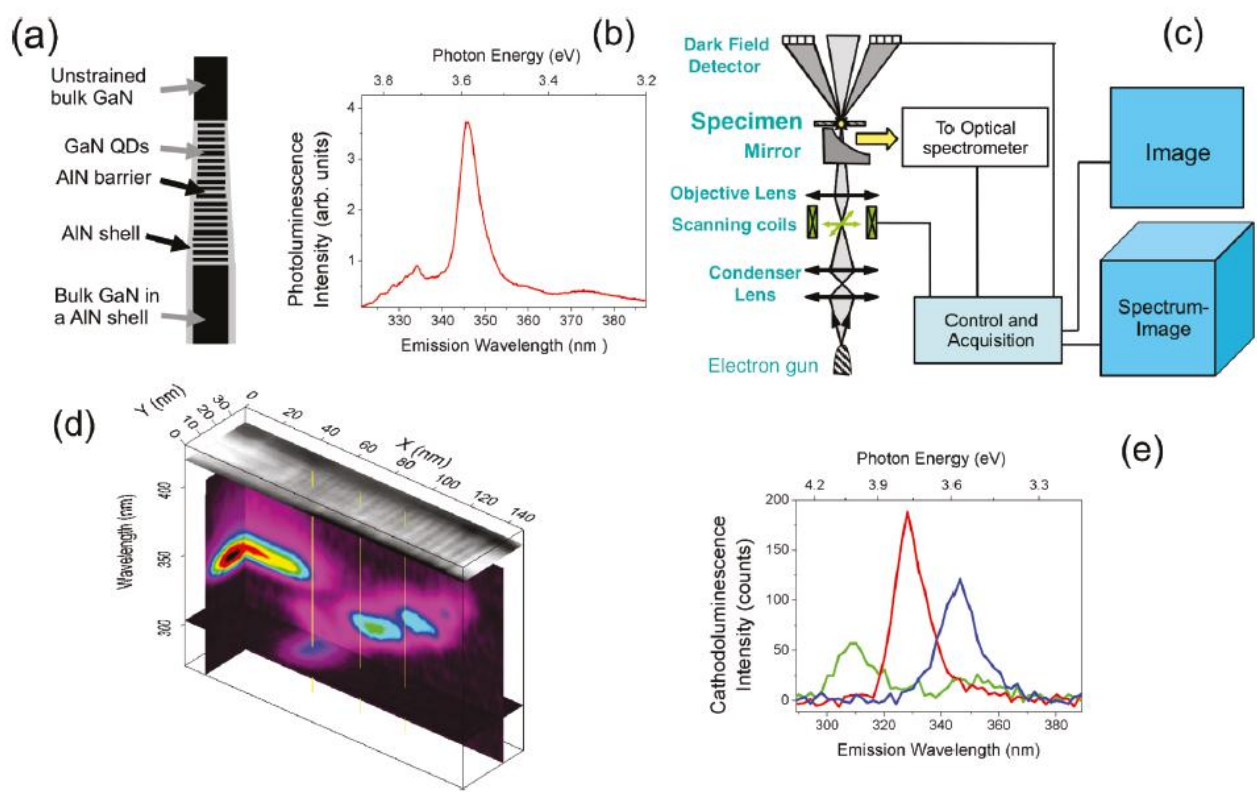

**Figure 1.** (a) Scheme of a GaN/AlN QDiscs stack in a nanowire. (b) Photoluminescence of a typical GaN/AlN nanowire as schematized in (a) (T = 4.2K). The contributions of the different QDiscs constituting the nanowire cannot be disentangled. (c) Principle of the cathodoluminescence spectral imaging setup in a STEM. (d) Typical output of a spectral imaging experiment. The top image is the HAADF image (GaN appears whiter, AlN darker), acquired simultaneously with the spectrum image. At each pixel (ca. 1 nm2) a cathodoluminescence spectrum is acquired. After acquisition of the whole data set, the information can be retrieved along various planes of the position-spectral 3D space. (e) Cathodoluminescence spectra extracted from the previous data set at position indicated by the thin yellow cylinders in (d). The contributions of different QDiscs are much better disentangled.

Quantum confinement effects have been studied on GaN/AlN nanowires (NW) containing a series of 20 GaN/AlN Si doped QDiscs embedded in the middle of the NW (see see Figure 1a). The NW were grown by using radio frequency plasma-assisted molecular beam epitaxy on Si (111) substrates under N-rich atmosphere at 790 °C (see [22,23] for growth details). The NWs have a wurtzite structure and grow along the c-axis. They are approximately 1-1.2 $\mu$ m long and have a diameter between 80 nm and 20 nm. The GaN QDiscs are doped with Si at a nominal concentration of $2.10^{19}$cm$^{-3}$. They have a thickness typically in the range from 1 nm (4 ML) to 3.4 nm (13 ML) and are separated by ca 3 nm thick AlN barriers. The disc thickness progressively increases towards the NW top. The QDiscs are packed between two GaN ends of approximately equal length, the lower one being totally surrounded by an AlN shell while the upper one is left free of coverage, as depicted in Figure 1a. The shell appears during the deposition of AlN barriers because AlN easily nucleates on the nanowire lateral facets. [23] The shell is around 5-10 nm thick in the lower NW part and its thickness progressively decreases in the QDisc region. While the QDisc thickness is smaller or equal to the Bohr radius of the exciton in GaN, [24] their diameter is substantially higher (10 nm), resulting in a strong quantum confinement along the growth axis, while the in-plane confinement is almost negligible. These GaN/AlN NWs have already demonstrated their potential for visible-blind nano-photodetectors. [25] However, their optical spectra are intricate and the attribution of spectral features can be ambiguous (see Figure 1b), therefore the ultimate spatial and spectral resolution is required to monitor their properties.

We developed a dedicated Cathodoluminescence (CL) system that fits a regular Scanning Transmission Electron Microscope (STEM) Vacuum Generators HB-501. The system is optimized to have a high collection angle (1.4 $\pi$ srd), yet keeping a good spectral resolution (0.3 nm) with a minimized signal loss in transmission optics. This, combined with the high brightness of the cold field emission gun, allows for high spatial (ca 1 nm) and spectral resolution within low acquisition times ranging typically from 20 to 500 ms per spectrum. In this STEM, an electron source focused on a sample interacts with it both quasi-elastically (giving rise to the High Angle Annular Dark Field - HAADF- signal) and inelastically (giving rise to the CL signal) (see Figure 1c and supple-mentary information). When the nanometric beam scans the sample, both an HAADF image, and a 3D CL spectrum-image (SPIM) are recorded in parallel (see Figure 1d). Hence each pixel (typi-cally 1 nm$^2$) of the HAADF image has a corresponding spectrum, as exemplified in Figure 1d and e. This makes the structural/optical properties relationship straightforward. The STEM was op-erated at 60 kV, which effectively reduced visible beam damages while still providing high probe currents and good spatial resolution. The sample was maintained at 150 K giving an increased emission efficiency and sharper spectral information than at room temperature. We a posteriori obtained high resolution HAADF images in a NION USTEM to get the exact number of atomic planes (see supplementary information) constituting the same individual NWs.

As an illustration, Figure 2 presents maps where, alongside the HAADF image acquired si-multaneously to the CL, the main optical properties are plotted. As several peaks may arise at a single probe position (see Figure 1e), for the sake of simplicity, we only map the principal optical properties (energy peak position, intensity, Full Width at Half Maximum, FWHM) of the most intense peak in the spectrum of a given pixel (see supplementary information). It is seen that the GaN end surrounded by the AlN shell (left part) is emitting strongly at $\lambda_{GaN/AlN}$ = 346.6 nm, at slightly lower wavelength than the bulk GaN ($\lambda_{GaN}$ = 359.5 nm). The value of $\lambda_{GaN}$ (measured on an emitting AlN-free side of another NW) is in close agreement with the value reported in the literature for the Near Band Edge (NBE) line at this temperature both in GaN epitaxial layers [26] and in binary GaN nanowire ensembles. [27] The slight blue-shift of the NBE in the GaN end surrounded by AlN is attributed to the compressive strain [28] exerted by the AlN shell. In the QDisc region two groups of optical signals are present, featuring lower (resp higher) emission wavelengths than that of the GaN lower end. Indeed, the QDisc thickness varies along the NW with some dispersion within a trend of increase from the beginning to the end of the growth, as observed in Figure 2a. Therefore, the general tendency is that as the QDisc thickness increases so does its emission wave-length. For the last QDiscs and the pure GaN ends, the emission intensity is vanishingly low. This is most likely due to the fact that the AlN shell passivates the surface states of GaN responsible for non-radiative recombinations and prevents carriers from reaching the NW surface. In some parts of the maps (two of them emphasized by black dashed rectangles) unique optical features such as constant emitting wavelength (Figure 2c) or peaked maxima of intensity (Figure 2d) can be related to a unique QDisc identified on the HAADF acquired in parallel. The optical signal is localized close to individual QDiscs, both along and perpendicular to the NW direction. When the electron probe moves away from a QDisc, the emission wavelength is kept constant while the intensity drops drastically in both directions. Also, no specific signal can be associated to the lateral side of the NW, asserting that the spectral features different from the GaN bulk emission can be asso-ciated solely to QDisc and not, e.g., to defects at the NW surface. We note that this 2D mapping of few nanometers QDiscs embedded in a hundred nanometers long NW can be obtained in only a few minutes. However, despite the very high spatial resolution obtained in the 2D mapping the attribution of unique spectral features to unique QDiscs remains ambiguous if relying only on data such as displayed on Figure 2.

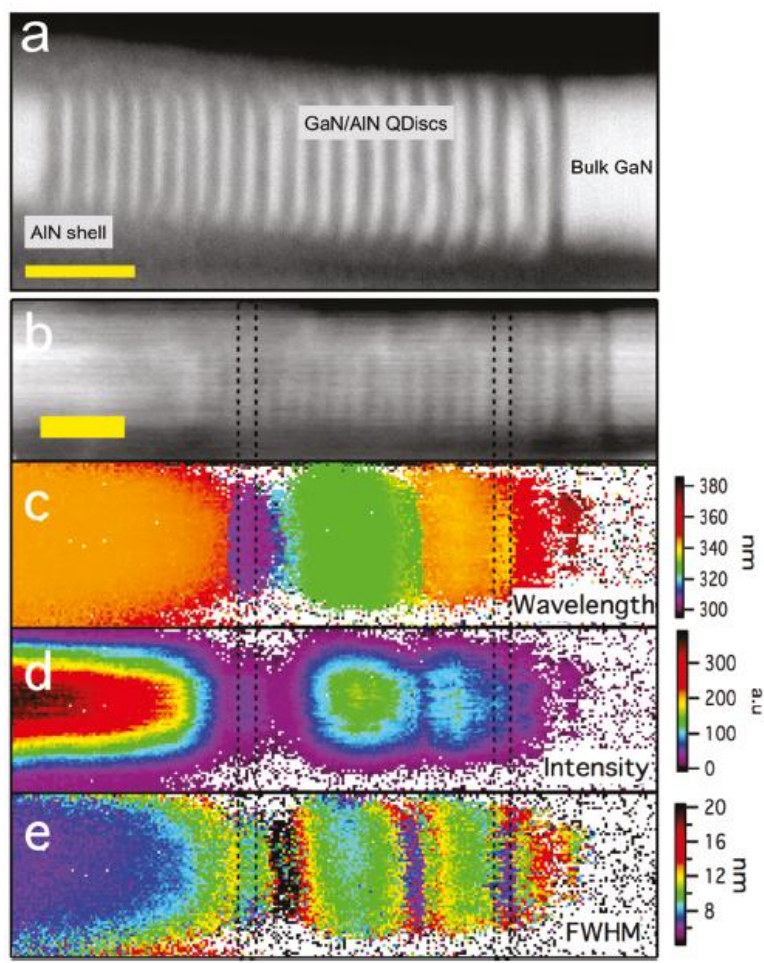

**Figure 2.** Spectral imaging of a GaN/AlN NW with 20 QDisks. It wasacquired with 256 per 64 pixels of spatial sampling, 0.6 nm/pixel, and with 256 pixels of spectral sampling, 2 nm/pixel (wavelength), within less than 6 min. (a) HAADF image of a GaN/AlN nanowire (GaN appears whiter, AlN darker). Scale bar is 20 nm. (b) HAADF of the NW, acquired simultaneously with the CL. Scale bar is 20 nm. (c) Wavelength position of the most intense peak. (d) Intensity of the most intense peak. (e) Full width at half-maximum of the most intense peak. Some individual spectral features can be correlated to some individual quantum disks, as emphasized by the black dashed rectangles.

In order to clearly identify spectral features of individual QDiscs, we systematically performed spectral imaging in a line along the NWs with a better spectral resolution and represented it as a 2D map combining spectral and spatial information. This is shown for another NW in Figure 3 (see also Supplementary Figures 2,3,4 and 6 ). There, the vertical axis indicates the emission wavelength and the horizontal axis indicates the position along the NW, which can be easily related to its structure through the HAADF profile on top. At a glance, each spectral feature is perfectly discriminated in the 2D map (see also the enlarged part in the inset) because it clearly peaks in the combined spatial/spectral space at a well defined wavelength and at a position corresponding to a QDisc (determined by the HAADF profile). Despite the fact that different spectral signatures may overlap spatially (at a given beam position, different emission signals are triggered), in the spectral-spatial combined space, each QDisc is associated with a unique well-identified 2D maximum. In Figure 3, almost all QDiscs identified in the HAADF profile can be assigned to a spectral feature and each spectral feature can be assigned to a QDisc. Thus, we demonstrate here that optical features spectrally separated by 4 nm (i.e. 40 meV at $\lambda = 350$ nm) and arising from QDiscs whose centers are separated spatially by 6 nm can be straightforwardly discriminated. The achieved high spatial resolution (better than the spatial delocalization) and spectral resolution (better than the Full Width at Half Maximum) definitely demonstrates the potential of spectral imaging with nanometer spatial and nanometer (few tens of meV) spectral resolution. In fact, in order to gain signal to noise ratio, speed and spatial resolution, it is common place in CL experiments to perform panchromatic imaging (acquiring the integrated signal over the whole spectrum with a single broad band detector) or monochromatic imaging (when the signal is integrated over a short spectrum range centered on a feature of interest). Such signals are however ambiguous, as they peak in several QDiscs (see Supplementary Figure 1 for example). Similarly, without a small electron probe, the spectra of single QDiscs would be acquired without sufficient discrimination (see the discrimination between the spectra of two neighboring QDiscs in the right panel of Figure 3, as well as the broaden information obtained with a degraded spatial resolution).

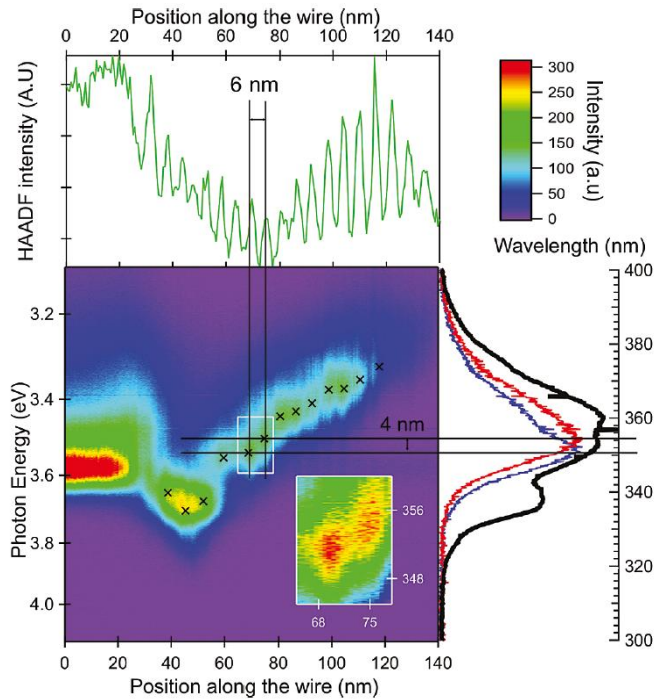

**Figure 3.** Spatial and spectral resolution of individual QDisks in spectral imaging. The central image shows the spectral imaging of a NW along a central line. Parameters are: step size = 0.6 nm, spectral resolution 0.36 nm, dwell time per spectrum 0.5 s. Emission from individual quantum disks is readily seen. Crosses indicate the maximum of the features fitted in this mixed spectral-spatial space (see Supporting Information). The top graph exhibits the HAADF profile (acquired simultaneously with the CL). The centers of GaN QDisks appear as peaks in the HAADF signal and are well correlated with maxima of CL emission in the spectralspatial map. Inset: Magnified view of the spectral image part enclosed in a white rectangle. Two QDisks signals can be straightforwardly discriminated although the QDisks centers are spatially separated by only 6 nm and spectrally separated by 4 nm (40 meV). Corresponding spectra are displayed on the right. A reconstructed spectrum is also shown in black to illustrate the badly resolved emission that would be obtained by integrating the signal on a bigger (50 nm) spatial range (position 46-96 nm).

We note that the signal delocalization, defined as the spatial spreading of the intensity QDisc in the combined spatial/spectral space, has a small extent (see Figure 3 and Supplementary Figure 6), much smaller than the diffusion length of excited carriers in AlN (about 100 nm [29]) and in GaN (about 120 nm [30]). This results from both the high recombination efficiency in the QDiscs and the large band gap contrast between GaN and AlN helping to efficiently capture and localize generated carriers. We expect such enhanced signal localization to occur in any nanostruc-tured material, giving the opportunity for such nanocathodoluminescence to have a broad range of applications in nanotechnology related materials as compared to bulk materials.

Knowing that individual spectral features can be unambiguously attributed to individual QDiscs, and using High Resolution HAADF images acquired after the CL experiments on the same indi-vidual QDiscs, we are now in position to compare the optical properties of individual QDiscs with their atomic level dimension (number of monolayers, ML), the size of their immediate neighbor-hood (AlN barrier thickness) and their relative positions within the NW. The results are plotted on Figure 4. We first note that the attribution of emission lines to individual QDiscs as small as 1 nm (4 MLs) is made possible by our technique, and that more generally this represents, to the best of our knowledge the first direct comparison between the optical properties and the structure measured with ML resolution of the same QDiscs. The tendency is an increase of the emission wavelength with the number of MLs (Figure 4a)), as expected for a quantum confinement effect. However, two main effects have to be discussed in detail. 1. The energies of the large QDiscs are smaller than the gap of the bulk GaN and 2. for a given thickness of the QDisc, the energies are largely scattered. Point 1 is due to the so-called Quantum-Confined Stark Effect [7] (QCSE). In the QCSE, the bending of the electron and hole energy levels by applied or internal fields leads to a spatial separation of the electron and hole wavefunctions and to a reduction of the transition energy between them. The resulting red-shift of the emission line can be so large that, for a large enough QDisc (more than ca 2.6 nm thick in the present study), the transition energy can be smaller than the bulk gap, as clearly seen on several QDs (see eg Figure 2, Figure 3 and Supplementary Figures 2,3,4). Such a redshift cannot be explained by the Si doping into the GaN discs, as this would result in a blueshift of the lines. The internal field in the present heterostructure stems from the well documented pyro- and piezo-electricity in polar GaN and AlN. [31] The built in field experi-enced by a particular QDisc will depend on the exact size of the QDisc and of that of the AlN barriers. This is reproduced by the simulation shown on Figure 4b (crosses). The simulations are based on the self-consistent resolution of the one dimensional Schrödinger-Poisson equations using the effective mass approximation and assuming the doping in the QDiscs to be $2 \cdot 10^{19} CM^{-3}$ (as described in the supplementary information). The material parameters are taken from. [32] This simulation does not account for the strain relaxation: the GaN QDiscs are supposed to be relaxed and the AlN barriers are fully strained. A good agreement between the spectroscopic results and simulations is achieved for small QDiscs. For the large QDiscs the experimental data trend can be reproduced if a reduced value of the internal field is considered. The best fit is obtained for 4 MV/cm electric field discontinuity which corresponds to a zero piezoelectric contribution (shown in Figure 4b and Supplementary Figure 7). The reason for this reduction of the internal field in comparison to the GaN/AlN quantum well case is not fully understood and requires further exper-imental and theoretical work. It could be related to the strain relaxation (not taken into account in the present simulation) and to the electric field screening with carriers. Indeed, the effective carrier concentration in the QDiscs is unknown and may differ from the nominal doping. It should be noted that the trend of having a weaker QCSE in NWs than in quantum wells has also been observed in recent experimental works on the GaN/AlN QDisc photoluminescence [8] and studied by more advanced theoretical calculations for isolated QDisc close to the NW top. [21] To the best of our knowledge, our results are the first demonstration of QCSE on individual QE with the size and environment known with atomic resolution, and should serve as validation of state-of-the-art theories. [21]

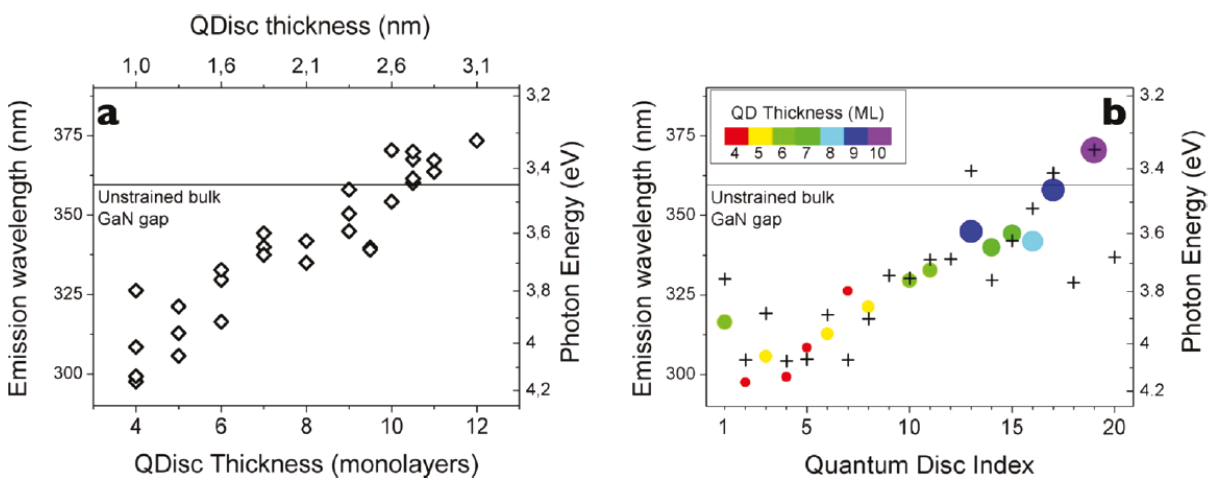

**Figure 4.** (a) CL energy as a function of QDisks thickness measured with monolayer resolution of the individual QDisks in the NWs of Figure 2 and Figure 3. For a given thickness, the dispersion in emission wavelength values is caused by varying positions within the NWs. (b) CL energy of individual QDisks shown as they are distributed inside a singleNW(same NWas that of Figure 2). The experimental values are given by full symbols whose color and size indicate the thickness in MLs of each QDisk according to the legend. The results of the simulations in the assumption that only a pyroelectric field is present are given by the black crosses. The emission of each QDisk is clearly affected not only by its size but also by its exact position along the NW. For a given QDisk size (same marker size and color) the emission wavelength systematically increases from the left to the right of the NW.

For a constant QDisc thickness and identical fields, the red shift due to the QCSE should be constant. However, in the present nanostructures, fields depend on the exact dimensions of the QDiscs and of their AlN barriers as well as the local strain. [21] Otherwise speaking, the apparent scattering of the energy value for a given number of monolayer (point 2) may find several reasons and is best understood by considering the energy dispersion of the QDiscs as a function of theirindex (position along the NW, starting from the AlN-covered GaN end), as presented as an example for the NW of Figure 3 in Figure 4b. The scattering of energy for a given QDisc size predicted by the calculations is much smaller than experimentally observed. The origin of the scattering must be found elsewhere, most probably in the different strain state of the QDiscs located at different positions which changes the bandgap and shifts the emission lines. Indeed, the strain decreases from the low index (thick AlN shells) to the high index (thin AlN shells) and at the same time, the energy of the QDisc with the same number of MLs systematically decreases with increasing index (see Figure 4).

It is worth commenting on the potential perturbing effect of the electron beam on the emission properties of the QDs. Photoluminescence experiments on large GaN/AlN quantum dots have demonstrated that the internal electric field can be screened by photogenerated carriers even at a moderate optical excitation [33] resulting in a drastic blue-shift of the emission energy as a function of the excitation intensity. In our case, when the beam is moved few nanometers away (in either the radial or longitudinal direction) from a given quantum disc its luminescence intensity drops by an order of magnitude or more. Despite this intensity variation, no wavelength shift is measurable (see Figure 2 and Supplementary Figure 6). This advocates for the effect of the excitation by the electron beam to be linear, differently from previous CL measurements [16] at higher beam currents and lower electron energy (thus much larger cross section). This allows us to exclude the possibility of the electric field screening with generated carriers.

In conclusion, we have demonstrated two-dimensional Cathodoluminescence spectral imaging with nanometer spatial resolution and 10 meV spectral resolution on a device-oriented stack of quantum emitters built in a nanowire. With this technique we are able to unambiguously identify the origin of the observed spectral features. We determined the emission wavelength of individual GaN/AlN quantum discs whose thickness is known with monolayer resolution in combination with high resolution imaging. The direct correlation between the QDiscs size and emission wavelengths allowed us to evidence the quantum confined Stark effect leading to an emission below the band gap for discs thicker than 2.6 nm and to probe the magnitude of the internal electric field. We show that the internal electric field in the studied QDiscs is smaller than what is expected in the quantum well case. We also demonstrate the influence of the strain exercised by an AlN shell on the inten-sity and the emission energy of the nanowire and the QDiscs. In particular, we evidence the impact of the strain variation on the energy dispersion between QDiscs of identical size. The capabilities of the present technique open the way both for fundamental studies of quantum confinement in closely packed quantum emitters (for example, Stranski-Krastanow GaN/AlN or InAs/GaAs quan-tum dots) and for characterizations of optoelectronic devices based on nanostructured materials (for example, study of In segregation in InGaN/GaN light emitting diodes). It will also contribute to further refine the existing models for quantum confinement in these systems.

References

(1)   Wolf, S. A.; Awschalom, D. D.; Buhrman, R. A.; Daughton, J. M.; von Molnar, S.; Roukes, M. L.; Chtchelkanova, A. Y.; Treger, D. M. Science 2001, 294, 1488–1495.

(2) Morkoc, H.; Strite, S.; Gao, G. B.; Lin, M. E.; Sverdlov, B.; Burns, M. Journal Of Applied Physics 1994, 76, 1363–1398.

(3) Yan, R. X.; Gargas, D.; Yang, P. D. Nature Photonics 2009, 3, 569–576.

(4) Xu, S. J.; Chua, S. J.; Mei, T.; Wang, X. C.; Zhang, X. H.; Karunasiri, G.; Fan, W. J.; Wang, C. H.; Jiang, J.; Wang, S.; Xie, X. G. Applied Physics Letters 1998, 73, 3153–3155.

(5) Gisin, N.; Ribordy, G. G.; Tittel, W.; Zbinden, H. Reviews Of Modern Physics 2002, 74, 145–195.

(6) Machhadani, H. et al. New Journal Of Physics 2009, 11, 125023.

(7) Miller, D. A. B.; Chemla, D. S.; Damen, T. C.; Gossard, A. C.; Wiegmann, W.; Wood, T. H.; Burrus, C. A. Physical Review Letters 1984, 53, 2173–2176.

(8) Renard, J.; Songmuang, R.; Tourbot, G.; Bougerol, C.; Daudin, B.; Gayral, B. Physical Re-view B 2009, 80, 121305.

(9) Rigutti, L.; Tchernycheva, M.; Bugallo, A. D.; Jacopin, G.; Julien, F. H.; Furtmayr, F.; Stutz-mann, M.; Eickhoff, M.; Songmuang, R.; Fortuna, F. Physical Review B 2010, 81, 045411.

(10) Nelayah, J.; Kociak, M.; Stephan, O.; de Abajo, F. J. G.; Tence, M.; Henrard, L.; Taverna, D.; Pastoriza-Santos, I.; Liz-Marzan, L. M.; Colliex, C. Nature Physics 2007, 3, 348–353.

(11) Pennycock, S. J.; Howie, A. Philosophical Magazine A 1979, 41, 809–827.

(12) Garayt, J. P.; Gerard, J. M.; Enjalbert, F.; Ferlazzo, L.; Founta, S.; Martinez-Guerrero, E.; Rol, F.; Araujo, D.; Cox, R.; Daudin, B.; Gayral, B.; Dang, L. S.; Mariette, H. Physica E-Low-Dimensional Systems & Nanostructures 2005, 26, 203–206.

(13) Yamamoto, N.; Bhunia, S.; Watanabe, Y. Applied Physics Letters 2006, 88, 153106.

(14) Strunk, H. P.; Albrecht, M.; Scheel, H. Journal Of Microscopy-Oxford 2006, 224, 79–85.

(15) Yamamoto, N.; Araya, K.; de Abajo, F. J. G. Physical Review B 2001, 64, 205419.

(16) Jahn, U.; Ristic, J.; Calleja, E. Applied Physics Letters 2007, 90, 161117.

(17) Lim, S. K.; Brewster, M.; Qian, F.; Li, Y.; Lieber, C. M.; Gradecak, S. Nano Letters 2009, 9, 3940–3944.

(18) Duxbury, N.; Bangert, U.; Dawson, P.; Thrush, E. J.; Van der Stricht, W.; Jacobs, K.; Moer-man, I. Applied Physics Letters 2000, 76, 1600–1602.

(19) Smeeton, T. M.; Kappers, M. J.; Barnard, J. S.; Vickers, M. E.; Humphreys, C. J. Applied Physics Letters 2003, 83, 5419–5421.

(20) Galtrey, M. J.; Oliver, R. A.; Kappers, M. J.; Humphreys, C. J.; Stokes, D. J.; Clifton, P. H.; Cerezo, A. Applied Physics Letters 2007, 90, 061903.

(21) Mojica, D. C.; Niquet, Y. M. Physical Review B 2010, 81, 195313.

(22) Songmuang, R.; Landre, O.; Daudin, B. Applied Physics Letters 2007, 91, 251902.

(23) Tchernycheva, M.; Sartel, C.; Cirlin, G.; Travers, L.; Patriarche, G.; Harmand, J. C.; Dang, L. S.; Renard, J.; Gayral, B.; Nevou, L.; Julien, F. Nanotechnology 2007, 18, 385306.

(24) Kawakami, Y.; Peng, A. G.; Narukawa, Y.; Fujita, S.; Fujita, S.; Nakamura, S. Applied Physics Letters 1996, 69, 1414–1416.

(25) Rigutti, L.; Tchernycheva, M.; Bugallo, A. D.; Jacopin, G.; Julien, F. H.; Zagonel, L. F.; March, K.; Stephan, O.; Kociak, M.; Songmuang, R. Nano Letters 2010, 10, 2939–2943.

(26) Shan, W.; Schmidt, T. J.; Yang, X. H.; Hwang, S. J.; Song, J. J.; Goldenberg, B. Applied Physics Letters 1995, 66, 985–987.

(27) Furtmayr, F.; Vielemeyer, M.; Stutzmann, M.; Laufer, A.; Meyer, B. K.; Eickhoff, M. Journal Of Applied Physics 2008, 104, 074309.

(28) Chuang, S. L.; Chang, C. S. Phys. Rev. B 1996, 54, 2491–2504.

(29) Petersson, A.; Gustafsson, A.; Samuelson, L.; Tanaka, S.; Aoyagi, Y. Applied Physics Letters 1999, 74, 3513–3515.

(30) Duboz, J. Y.; Binet, F.; Dolfi, D.; Laurent, N.; Scholz, F.; Off, J.; Sohmer, A.; Briot, O.; Gil, B. Materials Science And Engineering B-Solid State Materials For Advanced Technology 1997, 50, 289–295.

(31) Bernardini, F.; Fiorentini, V.; Vanderbilt, D. Physical Review B 1997, 56, 10024–10027.

(32) Vurgaftman, I.; Meyer, J. R.; Tansu, N.; Mawst, L. J. Applied Physics Letters 2003, 83, 2742–2744.

(33) Lefebvre, P.; Gayral, B. Comptes Rendus Physique 2008, 9, 816–829.

Acknowledgments

We acknowledge useful discussions and support from O. Stéphan and C. Colliex. The authors acknowledge financial support from the European Union under the Framework 6 program under a contract for an Integrated Infrastructure Initiative. Reference 026019 ESTEEM. The research described here has been partly supported by Triangle de la physique contract 2009-066T-eLight.

Correspondence and requests for materials should be addressed to M. K. (email: kociak@lps.u-psud.fr).